\documentclass[nonacm,sigplan]{acmart}
\usepackage[]{hyperref}
\usepackage{diagbox}
\usepackage{bm}
\usepackage{epsfig}
\usepackage{algorithmic}
\usepackage{graphicx}
\usepackage{textcomp}
\usepackage{xcolor}
\usepackage{multirow}
\usepackage{tikz}
\newcommand{\tabincell}[2]{\begin{tabular}{@{}#1@{}}#2\end{tabular}}
\begin{document}

\pdfpagewidth=8.5in
\pdfpageheight=11in


\pagenumbering{arabic}

\begin{titlepage} 

	\centering 
	
	\scshape 
	
	\vspace*{\baselineskip} 
	
	
	\rule{\textwidth}{1.6pt}\vspace*{-\baselineskip}\vspace*{2pt} 
	\rule{\textwidth}{0.4pt} 
	
	\vspace{0.75\baselineskip} 
	
	{\LARGE Inference of Component Effect on System Performance} 
	
	\vspace{0.75\baselineskip} 
	
	\rule{\textwidth}{0.4pt}\vspace*{-\baselineskip}\vspace{3.2pt} 
	\rule{\textwidth}{1.6pt} 
	
	\vspace{2\baselineskip} 
	
	
	
	\vspace*{3\baselineskip} 
	
	
	Edited By
	
	\vspace{0.5\baselineskip} 
	
	{\scshape\Large Chenxi Wang\\ Lei Wang\\ Wanling Gao\\ Fanda Fan\\ Guoxin Kang\\ Hongxiao Li\\ Yuchen Su\\ Jianfeng Zhan\\}

	\vspace{0.5\baselineskip} 

	\vfill 
	
	
	\epsfig{file=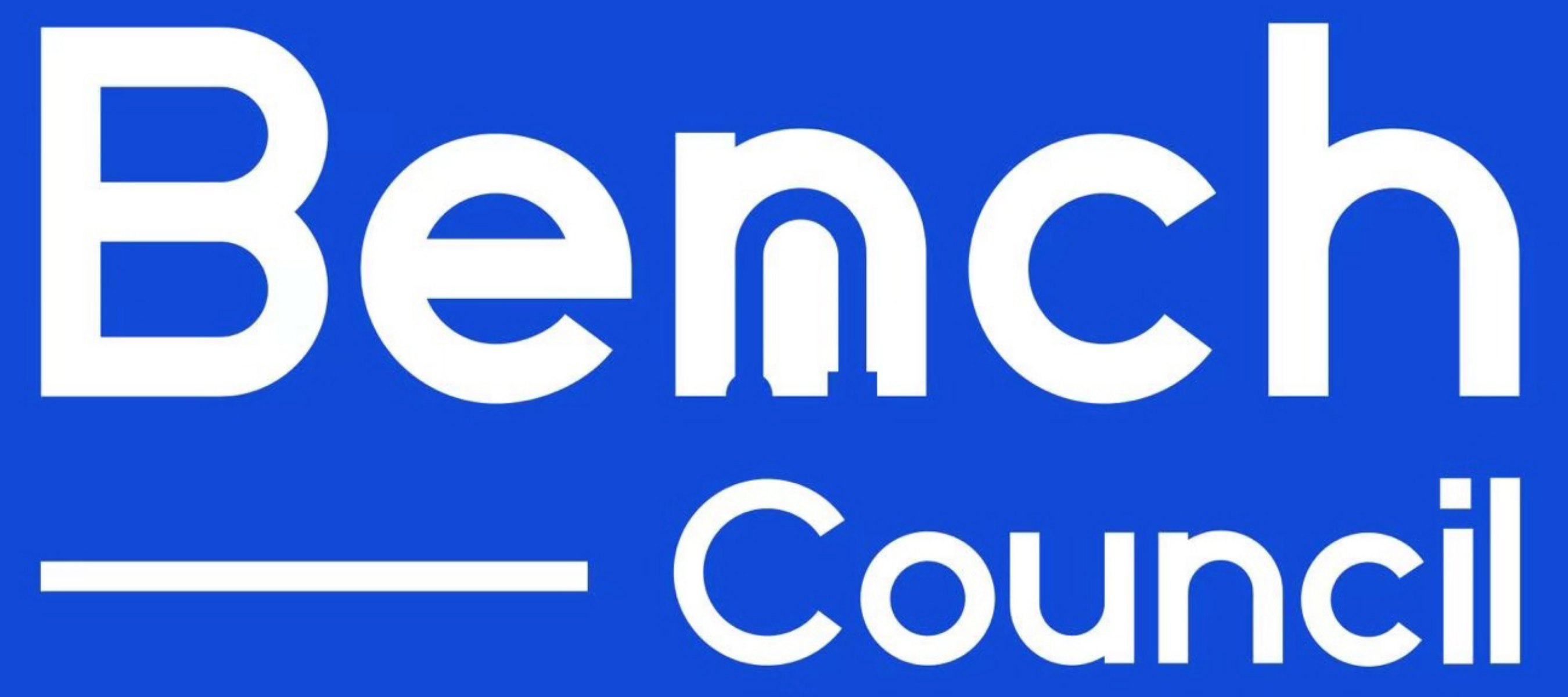,height=2cm}
	\textit{\\BenchCouncil: International Open Benchmark Council\\http://www.benchcouncil.org} 
	\vspace{5\baselineskip} 

	Technical Report No. BenchCouncil-CPU Evaluatology-2026 
	
	{\large June 23, 2026} 

\end{titlepage}

\title{Inference of Component Effect on System Performance}

\author{Chenxi Wang}
\affiliation{%
  \institution{Institute of Computing Technology, Chinese Academy of Sciences; State Key Lab of Processors, Institute of Computing Technology, Chinese Academy of Sciences; University of Chinese Academy of Sciences}
  \city{Beijing}
  \country{China}
}
\email{wangchenxi21s@ict.ac.cn}

\author{Lei Wang}
\affiliation{%
  \institution{The International Open Benchmark Council; Institute of Computing Technology, Chinese Academy of Sciences; University of Chinese Academy of Sciences}
  \city{Beijing}
  \country{China}
}
\email{wanglei_2011@ict.ac.cn}

\author{Wanling Gao}
\affiliation{%
  \institution{The International Open Benchmark Council; Institute of Computing Technology, Chinese Academy of Sciences; University of Chinese Academy of Sciences}
  \city{Beijing}
  \country{China}
}
\email{gaowanling@ict.ac.cn}

\author{Fanda Fan}
\affiliation{%
  \institution{The International Open Benchmark Council; Institute of Computing Technology, Chinese Academy of Sciences}
  \city{Beijing}
  \country{China}
}
\email{fanfanda@ict.ac.cn}

\author{Guoxin Kang}
\affiliation{%
  \institution{The International Open Benchmark Council; Institute of Computing Technology, Chinese Academy of Sciences}
  \city{Beijing}
  \country{China}
}
\email{kangguoxin@ict.ac.cn}

\author{Hongxiao Li}
\affiliation{%
  \institution{Institute of Computing Technology, Chinese Academy of Sciences; State Key Lab of Processors, Institute of Computing Technology, Chinese Academy of Sciences; University of Chinese Academy of Sciences}
  \city{Beijing}
  \country{China}
}
\email{lihongxiao19@mails.ucas.ac.cn}

\author{Yuchen Su}
\affiliation{%
  \institution{Institute of Computing Technology, Chinese Academy of Sciences;  University of Chinese Academy of Sciences}
  \city{Beijing}
  \country{China}
}
\email{suyuchen24s@ict.ac.cn}

\author{Jianfeng Zhan}
\authornote{Corresponding author.}
\affiliation{%
  \institution{The International Open Benchmark Council; Institute of Computing Technology, Chinese Academy of Sciences}
  \city{Beijing}
  \country{China}
}
\email{zhanjianfeng@ict.ac.cn}



\begin{abstract}

In a computer system, multiple components—such as the CPU, memory, and others—work together as a system whose performance can be directly measured. However, the effect of a component under investigation (CUI), e.g., CPU, on system performance cannot be directly measured and can only be inferred. Accurately inferring CUI effect on system performance is a critical issue. Our experiments reveal that the general-purpose rigorous methodologies, like Design of Experiments (DoE), Randomized Controlled Trials (RCTs), and a single-purpose empirical methodology, like SPEC CPU2017, can not address this issue effectively and efficiently.

We propose a rigorous methodology to address this issue: First, we identify a self-contained system (SCS) under the context of which we can completely understand how CUI and other essential components affect the system performance, and then we use a structural causal model methodology to represent and infer the causal effect of CUI on the system performance. 
We utilize this methodology and verify its correctness in the context of CPU design and evaluation.
Through theoretical analysis and pioneering controlled experiments, we systematically compare our methodology against three established methodologies: 
SPEC CPU2017, DoE, and RCTs. The results show that our methodology can achieve its goal effectively and efficiently, whereas others exhibit inherent limitations.

\end{abstract}

\maketitle
\thispagestyle{plain}
\pagestyle{plain}

\section{Introduction}\label{sec:introduction}

A computer system consists of several indispensable components, e.g., the CPU, OS, and memory, which work together as a system. The performance of a computer system can be directly measured. However, the effect of a component under investigation (CUI), e.g., the CPU, on system performance cannot be measured directly and can only be inferred.

In this context, the challenges lie in how to accurately infer the CUI effect on the system performance. If we wrongly infer the effect of a new design of a CPU due to the confounding biases from other essential components (Figure 1), we can not prove the effectiveness of our new design and may draw wrong conclusions. This issue is also fundamental in other engineering and science fields, as we build systems from components in almost all fields.

\begin{figure}[ht]
\centering
\includegraphics[scale=0.5]{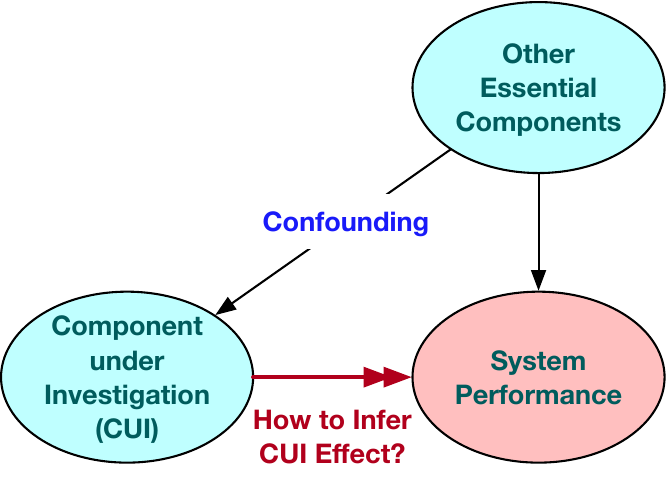}
\caption{The challenge of system design and evaluation lies in how to infer the CUI effect on the system performance, which is confounded by other essential components. The system performance can be directly measured, but the CUI effect on system performance can not be measured and can only be inferred.}
\label{fig1-com}
\end{figure}

We take CPU design and evaluation as a case study to verify different methodologies that can be used to infer the effect of a CPU (Figure~\ref{fig1-com}). For other essential components except workload, single-purpose empirical methodology like SPEC CPU2017 either fixes them to an empirical configuration or leaves them entirely unconstrained. Consequently, these components act as confounding when inferring the true CPU effect (Figure~\ref{fig1-com}(b)). Furthermore,
general-purpose rigorous methodologies like Randomized Control Trials (RCTs)~\cite{stolberg2004randomized,pearl2018book} and Design of Experiments (DoE)~\cite{jain1991art,telford2007brief,kirk2009experimental,fisher1971design,fisher1970statistical} can not effectively and efficiently address this problem. RCTs infer CPU effect by  random assigning system configurations into the treatment group CPU against the control group CPU. This random assignment inherently demands a colossal sample size to ensure high determinism 
in the derived causal effect (Figure~\ref{fig1-com}(c)). Conversely, DoE infer CPU effect through the exhaustive traversal of the entire combinatorial configuration space (Figure~\ref{fig1-com}(d)).

Our experiments reveal that the industry-standard CPU benchmark, SPEC CPU2017, suffered from a significant flaw. SPEC CPU just reports the overall system performance, which is contributed by all components, while overlooking the issue of how to infer the CPU effect on the system performance. Even more concerningly, it provided a performance score for the identical CPU with significant fluctuations (from 12.16\% to 436.80\% in our experiments), which violates an intuitive test oracle that the identical CPU should have the same performance score.
Worse still, when comparing different CPUs, changing standard compiler flags (\texttt{-O1}, \texttt{-O2}, \texttt{-O3}) consistently often induce complete performance rank reversals, which violate another test oracle that we derive using a structural causal model methodology.

Someone may argue that a rated performance number such as FLOPS suffices for CPU evaluation. However, extensive prior work has shown that peak performance numbers fail to reflect real-world system performance~\cite{panda2018wait, hoste2006comparing, hoste2007analyzing, hennessy2012computer, hennessy2019new, patterson2009computer}, a limitation also observed in our experiments, where such numbers do not reliably reflect system-level performance.

\begin{figure}[ht]
\centering
\includegraphics[scale=0.38]{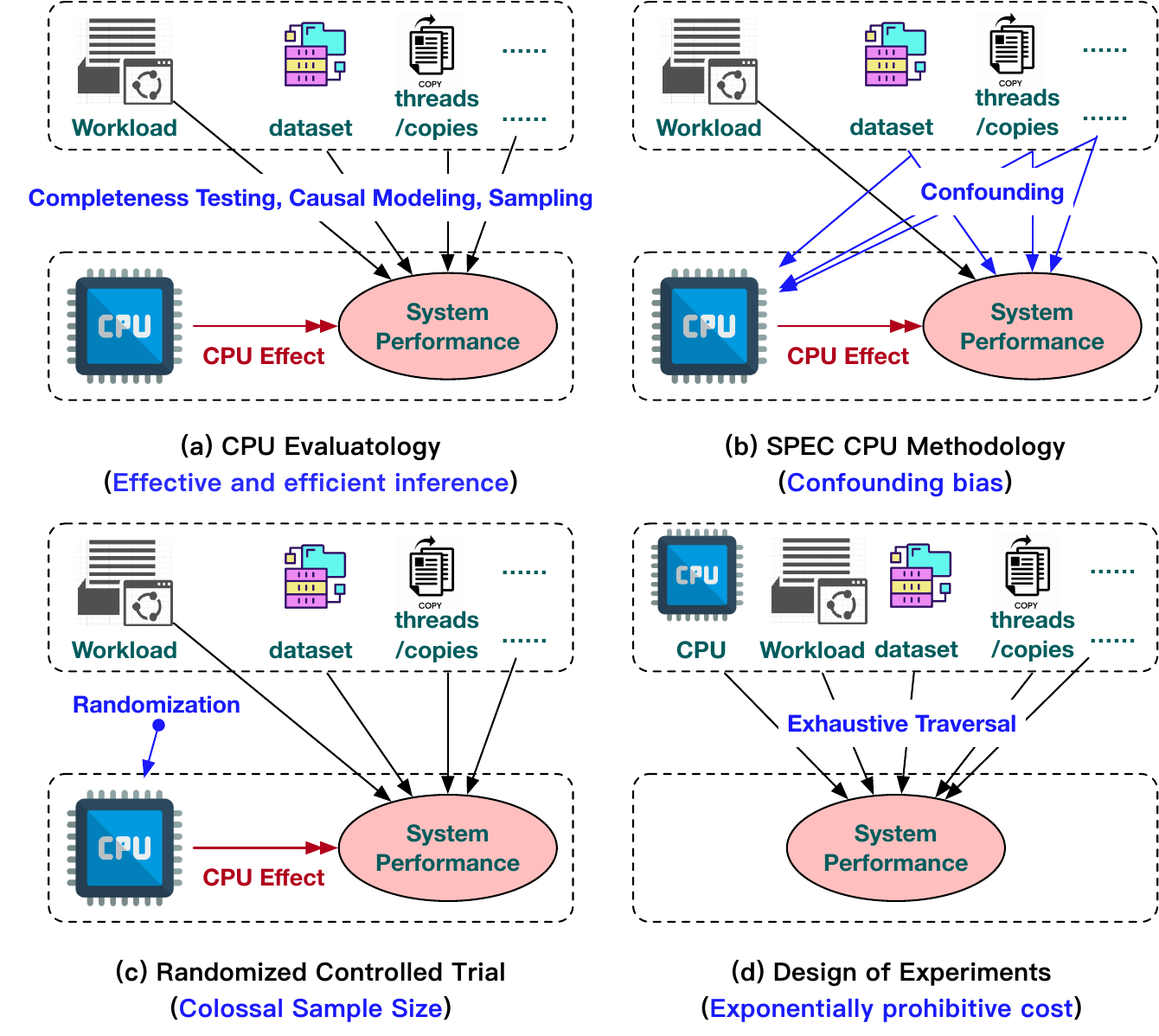}
\caption{
Methodology Comparisons of CPU Evaluatology, SPEC CPU, Randomized Controlled Trial, and Design of Experiments from a Perspective of Causal Effect.} 
\label{fig1-com}
\end{figure}

We propose a rigorous general-purpose methodology to address the above issue.  Our methodology is inspired by a general-purpose evaluation theory named Evaluatology~\cite{zhan2024evaluatology, zhan2025evaluatology}.  
We reframe the issue of inferring the CUI effect on the system performance as a causal inference problem. Firstly, for the CUI, we identify a self-contained system (SCS) under the context of which we can understand how the CUI and other essential components affect the system performance.  We validate the completeness of the SCS using a design-of-experiment (DoE) methodology---an F-test.  Finally,  we use the structural causal model methodology to infer the causal effect of CUI on the system performance. 

We embodied this methodology in the context of CPU evaluation and design, which we call CPU Evaluatology. An SCS is identified via F-test-based completeness testing. A structural causal model  (SCM) methodology is used to model how the CPU affects the system performance and to infer the true CPU effect under the backdoor criterion. Finally, random sampling is used to enable highly cost-effective inference.

We formalize two fundamental \textit{Test Oracles}—essentially architectural invariants—to validate the correctness of the inferred CPU effect using CPU Evaluatology: (1) The true performance delta of an identical CPU should be negligible; and (2) The cross-CPU performance difference (i.e., the true architectural advantage) under a uniform compiler flag should remain invariant. Guided by these oracles, we instantiate our methodology into an open-source validation tool, the CPUOracle suite.

To demonstrate the potential of our methodology, we infer the effect of a module within the CPU, Hyper-Threading (HT). Traditional static SPEC measurement misleads architects by reporting a deceptive \textbf{10\% performance penalty} on the Intel Xeon Gold 5120T upon HT activation. In contrast, rigorous methods, including our CPU evaluatology, DoE, RCTs, consistently uncover that HT incurs no such degradation and instead delivers a \textbf{marginal performance uplift of $\sim$2\%}, invariant across compiler flags. Furthermore, stratifying the outcomes by thread count reveals that the true HT effect on the Intel Xeon Gold 5120T is highly thread-dependent. For SPECfpspeed workloads, enabling HT degrades performance by 6.54\%–13.48\% when thread counts are bounded by the physical core limit. Conversely, it delivers an 8.57\%–21.56\% performance boost when thread counts lie between the physical and logical cores. Once the thread count surpasses the logical core capacity, enabling the HT yields no pronounced performance difference (fluctuations $\textless$ 2.5\%). Finally, CPU Evaluatology achieves this while shrinking the prohibitive estimation costs of DoE and RCTs by at least \textbf{two orders of magnitude}.

Our contributions are summarized as follows.

\begin{enumerate}
\item \textbf{Problem Formalization:} 
We recast the issue of inferring component effect on system performance as a causal inference problem and show that existing benchmark-driven methodologies fundamentally fail to isolate other components' effect, which are confounders.
\item \textbf{The methodology:} 
We propose a general-purpose rigorous methodology for inferring the effect of a component on the system performance and embody it in the context of CPU design and evaluation, which we call CPU Evaluatology.
\item \textbf{Verifying CPU Evaluatology:} 
We formalize two architectural invariants as test oracles and realize them in an open-source validation tool, CPUOracle suite, enabling rigorous verification of CPU performance inference.
\item \textbf{Comprehensive Case Study:} 
Using Hyper-Threading (HT) technology as a case study, we show that CPU Evaluatology effectively and efficiently addresses the issues of inferring the HT effect on system performance, where SPEC, DoE, and RCTs are either inaccurate or prohibitively expensive.
\end{enumerate}

The remainder of the paper is organized as follows. Section~\ref{pitfalls-spec} introduces our motivation. Section~\ref{Sec_M&R} discusses the background and related works. Section~\ref{Sec_Method} proposes our methodology, and Section~\ref{CPU_Eva_Method} applies it in CPU design and evaluation, and Section~\ref{Sec_EffectOracle} verifies its correctness. Section~\ref{TheEvaluationsection} shows the evaluation. Section~\ref{Sec_Conclusion} concludes. 

\section{Motivation}\label{pitfalls-spec}


SPEC CPU2017 is the industry standard for evaluating CPU. However, its metric, a SPEC score, essentially reflects the system performance that is contributed by all components of the system. Most importantly of all, it fails to accurately infer the CPU effect on the system performance.

\begin{table}
\centering
\caption{Machine information in our experiments. 
}
\resizebox{\columnwidth}{!}{
\begin{tabular}{llll}
\hline
\textbf{Machine}  & \textbf{CPU A}            & \textbf{CPU B}  & \textbf{CPU C}       \\ \hline
CPU Model Name & \textit{Intel Xeon Gold 5120T} & \textit{Intel Xeon Gold 5120T} &  \textit{Kunpeng 920} \\
Micro-architecture                &   \textit{Skylake}       &  \textit{Skylake}       & \textit{Taishan 110}                \\
ISA                &   \textit{x86-64}       &  \textit{x86-64}       & \textit{AArch64}                \\
Socket(s)          & 2  & 2                           & 2                          \\
Core(s) Per Socket & \textit{14} & \textit{14}                            & \textit{32}                         \\
Thread(s) Per Core & \textit{2}  & \textit{1}                           & \textit{1}                          \\
Compiler           & GCC9 & GCC9                    & GCC9                  \\ 
Memory             & 384GiB  & 384GiB             & 384GiB                     \\
Disk               & 2TiB & 2TiB   & 2TiB                       \\
Operating System   & \textit{Ubuntu 20.04.6 LTS}         & \textit{Ubuntu 20.04.6 LTS}   & \textit{Ubuntu 22.04.4 LTS}         \\
Linux kernel   & \textit{5.4.0-216-generic}         & \textit{5.4.0-216-generic}    & \textit{6.5.0-15-generic}          \\
\hline
\end{tabular}
}
\label{Machine-Inf}
\end{table}

From a perspective of structural causal models (SCMs), we reveal why the SPEC CPU methodology will introduce confounding bias. As shown in Figure~\ref{fig1-com}(b), other essential components like workload, dataset, and thread act as confounders in inferring the true CPU effect, as they dictate observed performance directly while simultaneously changing CPU execution state. SCMs eliminate confounding bias by exhaustively traversing all confounder distributions, SPEC CPU restricts this traversal solely to workloads. The other components are either statically fixed or left 
unconstrained. As a result, SPEC fundamentally fails to eliminate the confounding bias injected by any non-workload components. 

Furthermore, to quantitatively expose this vulnerability, we strictly adhered to the SPEC methodology (static workloads, datasets, compiler flags, copies/threads) to evaluate three processors (Table~\ref{Machine-Inf}): two x86-based Intel Xeon Gold 5120T CPUs (CPU A with HT enabled, CPU B disabled) and an ARM-based Kunpeng 920 (CPU C). Our profiling exposes systemic fragilities across two critical dimensions: identical-CPU volatility and cross-CPU inconsistencies.

\noindent\textbf{Observation I: Single-CPU Volatility Eclipses True CPU Effect.} 
Current benchmarking practices often fix the CPU and its workload configurations but leave the broader system components underspecified. We evaluated CPUs A and C across a spectrum of standard, real-world configurations, varying compilers (GCC, LLVM-Clang~\cite{lattner2008llvm}, ICC~\cite{ICC}), OS kernel versions (v3.10.0 to v6.5.0), memory capacities (64\,GiB to 768\,GiB), NUMA policies (NUMA aware and NUMA unaware), and disk configurations (800\,GiB to 2\,TiB). 

\begin{figure}[ht]
\centering
\includegraphics[scale=0.45]{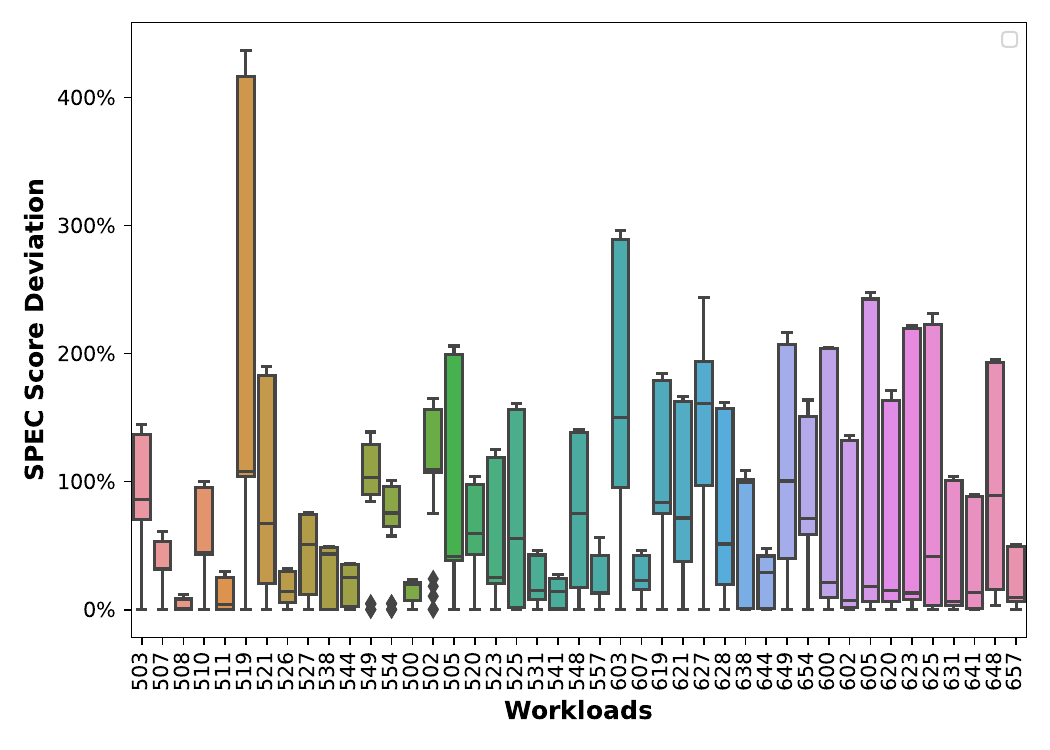}
\caption{
Using SPEC CPU2017 workloads and the \texttt{ref} dataset, the SPEC scores (overall performance) significantly differ when evaluating CPU A. 
We collected the experiment data from our experiments and the official release of SPEC CPU2017~\cite{CPU2017_results}. We showed the box plot~\cite{williamson1989box} of the deviation. 
}
\label{mot-dif}
\end{figure}

As illustrated in Figure~\ref{mot-dif}, for the single CPU (CPU A), the SPEC score for a single workload fluctuates wildly, ranging from 12.16\% to a staggering 436.80\% under different system configurations. Discrepancies at the sub-suite level are equally alarming: 84.54\% in \texttt{SPECfprate}, 88.13\% in \texttt{SPECint-}\\ \texttt{rate}, 119.27\% in \texttt{SPECfpspeed}, and 144.46\% in \texttt{SPECintspeed}. 
To rigorously quantify this, we performed an Analysis of Variance (ANOVA) and an F-test (Table~\ref{ANOVA-F-test-SPEC}) using CPUs and SPEC workload configurations as primary factors. The statistical outcomes are striking: systematic error accounts for up to 13.19\% of the total variance, completely eclipsing the CPU effect. At a 99\% confidence level, the effect of the CPU on the system performance is rendered statistically insignificant compared to the uncontrolled 
residual errors.

\begin{table}[htbp]
  \centering
  \caption{Analysis of Variation (ANOVA)  and F-test~\cite{jain1991art} for evaluating CPUs A and C using SPEC CPU2017. F-Computed are the computed F-values derived by dividing the component’s mean square by the mean square errors. F-Table are the critical F-values obtained via F-table lookup using p=0.01 and the corresponding degrees of freedom.
  }
  \resizebox{\columnwidth}{!}{
    \begin{tabular}{lcccc}
    \hline
    \textbf{Component} & \multicolumn{1}{l}{\textbf{Percentage of Variation}} & \multicolumn{1}{c}{\textbf{F-Computed}} & \multicolumn{1}{c}{\textbf{F-Table}} & \multicolumn{1}{c}{\textbf{F-Test (p < 0.01)}} \\\hline
    CPU(c) & 0.0944\% & 2.46  & 6.71 & \textit{Fail}  \\
    Workload(w) & 78.5722\% & 48.79  & 1.64 & Pass \\\hline
    c-w   & 8.1423\% & 5.06  & 1.64  & Pass \\\hline
    Errors & 13.1911\% &  \multicolumn{1}{c}{\enspace \enspace\diagbox{}{}} & \multicolumn{1}{c}{\enspace \enspace\diagbox{}{}} & \multicolumn{1}{c}{\enspace \enspace\diagbox{}{}} \\\hline
    \end{tabular}%
  }
  \label{ANOVA-F-test-SPEC}%
\end{table}%

\noindent\textbf{Observation II: Cross-CPU Evaluations are Fragile and Prone to Inversion.} 
Beyond single-CPU volatility, we demonstrate that cross-CPU comparisons (e.g., x86 vs. ARM) using standard SPEC metrics are highly brittle. Using the default machine configurations (Table~\ref{Machine-Inf}), we systematically varied compiler flags (\texttt{-O1}, \texttt{-O2}, \texttt{-O3}) and input datasets for the \texttt{SPECfpspeed} sub-suite.

\begin{table}[]
\centering
\caption{Floating-point performance of the three CPUs from Table~\ref{Machine-Inf}, evaluated via the \texttt{SPECfpspeed} sub-suite across various datasets and compiler flags. \texttt{ref} and \texttt{test} denote official SPEC datasets, whereas \texttt{ref+shift} and \texttt{test+shift} are synthetic datasets generated by injecting minor perturbations into the \texttt{ref} / \texttt{test} dataset. Performance ratios exceeding 1.0 are italicized. Minimum ratios across CPUs are underlined, and maximums are denoted by a superscript asterisk (*).}
\resizebox{\columnwidth}{!}{
\begin{tabular}{llcccc}
    \hline
    \multicolumn{1}{c}{\textbf{Performance ratios}} & \multicolumn{1}{c}{\textbf{}} & \texttt{ref} & \texttt{ref+shift} & \texttt{test} & \texttt{test+shift} \\ \hline
    \multirow{3}{*}{\textbf{CPU A / CPU B}}        & \texttt{-O1}                           & \underline{0.8632}       & $\textit{1.0708}^*$        & 0.9674              & 0.9595               \\
                                               & \texttt{-O2}                           & 0.8982       & \textit{1.0591}        & 0.9969               & 0.9529               \\
                                               & \texttt{-O3}                           & 0.8994       & \textit{1.0698}        & 0.9314               & \textit{1.0114}               \\ \hline
      \multirow{3}{*}{\textbf{CPU A / CPU C}}        & \texttt{-O1}                           & 0.9338       & 0.9119        & \underline{0.8003}               & 0.8396               \\
                                               & \texttt{-O2}                           & \textit{1.0431}       & \textit{1.0029}        & 0.9384               & 0.8571               \\
                                               & \texttt{-O3}                           & \textit{1.0249}       & \textit{1.0214}        & 0.8625               & $\textit{1.1388}^*$               \\ \hline
    \multirow{3}{*}{\textbf{CPU B / CPU C}}        & \texttt{-O1}                           & \textit{1.0818}       & 0.8516        & \underline{0.8272}              & 0.8750               \\
                                               & \texttt{-O2}                           & $\textit{1.1613}^*$       & 0.9469        & 0.9413               & 0.8995               \\
                                               & \texttt{-O3}                           & \textit{1.1396}       & 0.9547        & 0.9216               & \textit{1.1259}              \\ \hline
    \end{tabular}
}
\label{Mot-Cross-CPU}
\end{table}

As SPEC officially provisions a highly sparse input space (only \texttt{test}, \texttt{train}, and \texttt{ref} datasets), we synthesized over 30 valid datasets per workload that strictly preserve SPEC's official execution semantics (open-sourced in our CPUOracle suite). Table~\ref{Mot-Cross-CPU} captures the floating-point performance ratios between CPU pairs. Our analysis yields the following critical insights for existing benchmark-based architectural evaluation.

\textbf{Inadequacy of Static Configurations:} Employing static workload configurations is fundamentally insufficient to eliminate the confounding bias introduced by underlying system components. Treating the geometric-mean aggregated score—computed in strict adherence to SPEC's static-configuration mandate—as a proxy for true CPU performance suffers from severe volatility. Ratios between identical CPU pairs exhibit discrepancies exceeding 30\%, with configuration shifts capable of inducing complete rank reversals.

\textbf{SPEC Configurations Lack Representativeness:} The widely accepted SPEC baseline—using the \texttt{-O3} flag and the \texttt{ref} dataset—fails to capture the true architectural performance. Relying solely on this rigid configuration fundamentally misrepresents the true CPU performance. 

\section{Background and Related Work}\label{Sec_M&R}

This section maps the landscape of performance evaluation, highlights the gap our work bridges, and formalizes the theoretical framework of causal inference.

\subsection{Benchmarks, Measurements and Confounding Bias}\label{Sec-Bench-Pitfalls}

\textbf{Traditional Benchmarking Paradigms.} Computer architecture heavily relies on empirical methodologies, utilizing benchmark suites~\cite{panda2018wait,ferdman2012clearing,bienia2008parsec,bienia2008parsec1,woo1995splash,hoste2006comparing,hoste2007analyzing,hoste2006performance,hoste2007microarchitecture,shao2013isa,phansalkar2005measuring,phansalkar2007analysis,limaye2018workload,wang2023wpc,campanoni2010highly} (e.g., SPEC CPU~\cite{SPECCPU89,SPECCPU92,SPECCPU95,SPECCPU2000,SPECCPU2006,SPECCPU2017}, PARSEC~\cite{bienia2008parsec}, SPLASH~\cite{woo1995splash}) to proxy real-world performance~\cite{panda2018wait, ferdman2012clearing, shao2013isa, wang2023wpc}. The canonical approach aggregates execution outcomes across predefined, static system configurations to yield a synthesized metric (e.g., the SPEC score). 

\textbf{Measurement Error vs. Confounding Bias.} While benchmarks strive for standardizing workloads, researchers have long recognized that extraneous system configurations severely impact performance. Hennessy et al.~\cite{hennessy2019new} demonstrated that software-stack shifts (e.g., matrix multiplication algorithms) can induce performance gaps spanning orders of magnitude. Mytkowicz et al.~\cite{mytkowicz2009producing} exposed how innocuous environmental changes (e.g., link order, environment variables) cause pervasive measurement biases in architectural evaluation. Similarly, in the database domain, Wang et al.~\cite{wang2022study}, and Benson et al.~\cite{benson2024surprise} highlighted how experimental configurations dictate overarching performance conclusions. 
Crucially, prior works often treat these variations as mere ``measurement error'' to be averaged out. These findings are helpful to understand how different configurations impact the system performance, but they fail to formulate and address the issue of how to infer component effect on system performance under the condition that the effect can not be directly measured.

\subsection{Bridging Rigor and Intractability}\label{Sec-Rigor-Gap}

\textbf{The Intractability of Rigorous Methodologies.} 
Zhan et al.~\cite{zhan2024evaluatology,zhan2025evaluatology} recently proposed Evaluatology, formalizing evaluation as uncovering the effect of the evaluated object (EO) on the affected object (AO) and conducting deliberate experiments over well-defined configuration spaces.
In general science, disentangling effect relies on rigorous frameworks such as the Design of Experiments (DoE)~\cite{jain1991art,telford2007brief,kirk2009experimental,fisher1971design,fisher1970statistical}, Randomized Controlled Trials (RCTs)~\cite{stolberg2004randomized}, and do-calculus~\cite{pearl_causality_2000,pearl2018book}. In principle, applying DoE or RCTs directly to computer systems is practically intractable. Exhaustively traversing the massive, multi-dimensional configuration space (workload, dataset, compiler, OS, memory, disk) incurs prohibitive cost overheads.

Our contribution is to formulate a new problem of which several components work together as a system, and how to infer the causal effect of components on the system performance, which are overlooked by the previous causal inference work. Meanwhile, our methodology is innovative in terms that we combine three different methodologies: Evaluatology, causal structural model, and DoE to address the above issue.

\textbf{Statistical Approximations.} To navigate limited sample sizes, Chen et al.~\cite{chen2014statistical} proposed non-parametric statistical frameworks for comparing computer performance. Additionally, bootstrapping techniques~\cite{diciccio1996bootstrap} are proven to automatically construct highly accurate confidence intervals in complex probabilistic structures.
Inspired by these mathematical foundations and the formalization of Evaluatology, our work uniquely bridges the gap between rigorous causal inference and practical system evaluation. Instead of exhaustive traversal, we introduce a highly cost-effective stratified random sampling methodology (using workloads as strata) to infer the true CPU effect. This enables us to achieve the rigor of SCMs while reducing overhead by orders of magnitude.

\subsection{Causal Inference and Performance Modeling}\label{Sec-Causal}

\textbf{Utilizing Structural Causal Models in Systems.} Recent studies, such as those by Dave et al.~\cite{dave2023explainable} and Nasr-Esfahany et al.~\cite{nasr2025concorde}, utilize bottleneck or performance models to quantify hardware effect. Fundamentally, these models are domain-specific instantiations of causal graphs in computer systems. To formalize this, we leverage Structural Causal Models (SCMs)~\cite{pearl_causality_2000,imbens2015causal,pearl2018book} and Directed Acyclic Graphs (DAGs). A DAG establishes causal relationships via directed edges (e.g., Cause $X \rightarrow$ Effect $Y$) through three primitive junctions: chains ($A\rightarrow B\rightarrow C$), forks ($A\leftarrow B\rightarrow C$), and colliders ($A\rightarrow B\leftarrow C$). Within a DAG, \textbf{a causal path} constitutes a fully directed sequence wherein all constituent edges flow uniformly from $X$ to $Y$. In contrast, \textbf{a backdoor path} from 
$X$ (the root node) to $Y$
(the leaf node) is defined as any path for which there is constituent edge flow from $Y$ to $X$. In evaluating architectural hardware, the non-causal paths (backdoor paths)—introduced by other system components like compilers, OS, and memory—act as confounders. Accurately isolating the causal effect of $X$ on $Y$ inherently mandates the severance of all backdoor paths linking $X$
and $Y$ within the DAG. In the paradigm of SCMs, eliminating these backdoor paths entails an exhaustive traversal over the confounder distributions to rigorously control for confounding bias. To accommodate diverse structural scenarios for backdoor path obstruction, SCMs formalize robust mathematical frameworks, most notably the backdoor and frontdoor criteria. By modeling the system via DAGs, we can directly pinpoint the origins of confounding bias that compromise existing benchmarking methodologies.

\section{Methodology of Inferring CUI Effect}\label{Sec_Method}

This section presents our innovative methodology for inferring the CUI effect on system performance. 
Our approach is a general and rigorous framework consisting of two stages: first, to identify a SCS through completeness testing; and second, to utilize a structural causal model to represent and infer the causal effect of the component on the system performance. Functionally, the \textbf{\textit{input}} to our methodology consists of the CUI and the other essential components that together form the system under evaluation. Each component is associated with a specific configuration distribution. The \textbf{\textit{output}} is the inferred CUI effect.

\subsection{Identifying a self-contained system through completeness testing}\label{step-one}

The first stage 
identifies a \emph{self-contained system} (SCS) under the context of which we can understand how the CUI and other essential components affect the system performance.

A SCS has two distinguished characteristics. First, the system performance could be directly measured, acting as a ground truth. Second, it provides a sufficient boundary for understanding how the CUI and other essential components impact the system performance.

To ascertain the structural completeness of the SCS---effect- ively guaranteeing the determinism and convergence of the empirical measurements---we execute an F-test derived from the DoE framework. Treating the isolated components as categorical factors, the F-test quantifies their statistical significance relative to residual error introduced by uncontrolled systematic confounding, strictly operating at a 99\% confidence level (p$\textless$0.01). We consider a SCS complete if and only if all delineated components pass the F-test at the specified confidence level, thereby demonstrating statistical significance against the residual error and establishing a SCS as valid. Should any individual component fail to yield a statistically significant F-statistic, the SCS is inherently falsified as incomplete, mandating further granular decomposition of the system components. Through this procedure, the proposed methodology systematically identifies a SCS and rigorously validates its completeness.

\subsection{Representing and inferring the causal effect of the component within the self-contained system}~\label{step-two}
After identifying and validating an SCS,
we proceed to the second stage: 
representing the component within the SCS using a structural causal model and inferring its causal effect. This stage consists of two parts: modeling and inference. In the modeling part, we construct a structural causal model to formally characterize the causal relationships among the CUI, other components, and the overall system performance. In the inference part, we leverage the constructed model to estimate the true effect of the CUI.

\subsubsection{Modeling CUI evaluation causal graph}\label{step-two-1}
We formulate a causal graph within the structural causal model framework to characterize the intricate causal relationships among these components in CUI evaluation, thereby clarifying the causal origins of confounding bias. Note that the formulated causal graph is explicitly dedicated to CUI evaluation; accordingly, all modeled causal relationships are centered on the CUI. The central objective of CUI evaluation is to infer the true CUI effect, formalized by the isolated causal path: $CUI \rightarrow Performance$. Using the causal graph can explicitly expose the non-causal backdoor paths present in empirical measurements ($CUI \leftarrow X \rightarrow Performance$), thereby revealing how confounding bias arises and propagates, where $X$  denotes other essential components, which are the confounders.

\subsubsection{Inferring the true CUI effect}\label{step-two-2}

The accurate inference of the true CUI effect necessitates the systematic severance of all backdoor paths within the CUI evaluation causal graph. Guided by the do-calculus criterion, blocking these paths mandates strict control over all confounders, fundamentally requiring the exhaustive traversal of their permissible configuration spaces. We give Equation~\ref{EQ_CPU_effect}.

\begin{equation}
E(c_i) = \mathbb{E}_{X \sim P(X)} \left[ OE(c_i \mid X) \right],
\label{EQ_CPU_effect}
\end{equation}

where $c_i$ denotes the CUI, $X$  denotes other essential components (confounders), 
$P(X)$ is their distribution in the target population, $OE(c_i \mid X)$ is the overall system effect conditioned on $c_i$ and $X$, and $E(c_i)$ is the true effect of $c_i$.

While blocking backdoor paths through confounder control remains essential, the enormous combinatorial configuration space makes exhaustive traversal prohibitively expensive. We therefore redefine this space as a statistical population and treat the true CUI effect as its underlying population parameter. Based on this view, we adopt a stochastic sampling paradigm to estimate this parameter from samples, together with rigorous confidence levels and intervals, thereby enabling cost-effective inference of the CUI effect.

\section{Utilizing in CPU evaluation and design}\label{CPU_Eva_Method}

This section presents our innovative methodology in CPU evaluation and design, termed CPU Evaluatology.

\subsection{Identifying Key System Components through Completeness Testing}\label{step-one}

CPU Evaluatology targets to infer the effect of the CPU on the system performance. Roughly, we divide the self-contained system into a benchmark 
and the host machine (the underlying hardware platform and basic system software). 

Within this reference partition, the benchmark is delineated into four distinct components: workload, dataset, compiler flag, and copies/threads; concurrently, the host machine is decomposed into five components: compiler, OS, memory, disk, and CPU. 
We use the F-test to validate the completeness of the SCS. 

Leveraging the reference partitioning scheme of CPU Evaluatology, we executed the F-test on SPEC CPU2017 evaluations. In this setup, the benchmark-related components are configured with SPEC settings, while the host machine-related components excluding the CPU adopt the configurations listed in Table~\ref{Machine-Inf}. Collectively, we refer to these as the other essential component factor. As evidenced by Table~\ref{ANOVA-F-test-SPEC-EEC}, all structurally identified components successfully clear the F-test, exhibiting statistical significance against uncontrolled systematic confounding.

\begin{table}[htbp]
  \centering
  \caption{Analysis of Variation (ANOVA)  and F-test~\cite{jain1991art} for evaluating CPU A and CPU C using the system components of CPU Evaluatology (SPEC CPU2017 as workload). 
  }
  \resizebox{\columnwidth}{!}{
    \begin{tabular}{lcccc}
    \hline
    \textbf{Component} & \multicolumn{1}{l}{\textbf{Percentage of Variation}} & \multicolumn{1}{c}{\textbf{F-Computed}} & \multicolumn{1}{c}{\textbf{F-Table}} & \multicolumn{1}{c}{\textbf{F-Test (p < 0.01)}} \\\hline
    CPU(c) & 1.3770\% & 1577.66  & 6.78 & Pass  \\
    \tabincell{l}{Other essential\\~~components(o)} & 84.8895\% & 2315.69  & 1.70 & Pass \\\hline
    c-o   & 13.5833\% & 370.54  & 1.70 & Pass \\\hline
    Errors & 0.1501\% &  \multicolumn{1}{c}{\enspace \enspace\diagbox{}{}} & \multicolumn{1}{c}{\enspace \enspace\diagbox{}{}} & \multicolumn{1}{c}{\enspace \enspace\diagbox{}{}} \\\hline
    \end{tabular}%
}
  \label{ANOVA-F-test-SPEC-EEC}%
\end{table}%

Instead, revisiting Table~\ref{ANOVA-F-test-SPEC} in Section~\ref{pitfalls-spec}, we demonstrated that the canonical SPEC CPU2017 partitioning is inherently incomplete, as the inferred CPU effect failed the F-test, rendering it statistically insignificant against systematic confounding.

\subsection{Modeling causal graph of the CPU within SCS}~\label{step-two}

Once the key system components are structurally identified and their completeness rigorously tested, CPU Evaluatology instantiates a causal graph to model the intricate causal structures among these components in an SCS, thereby demystifying the causal origins of confounding bias.

\begin{figure}[ht]
\centering
\includegraphics[scale=0.57]{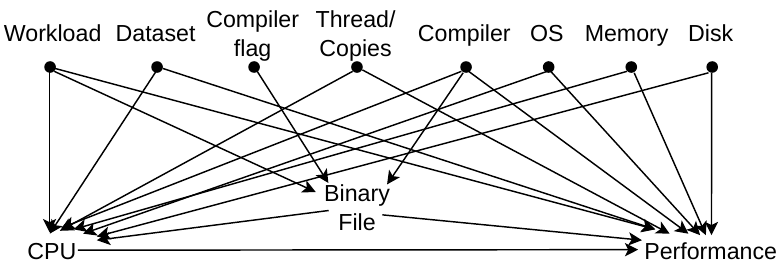}
\caption{
Causal graph of the SCS.} 
\label{fig-cpu-eva-causal}
\end{figure}

\begin{figure}[ht]
\centering
\includegraphics[scale=0.57]{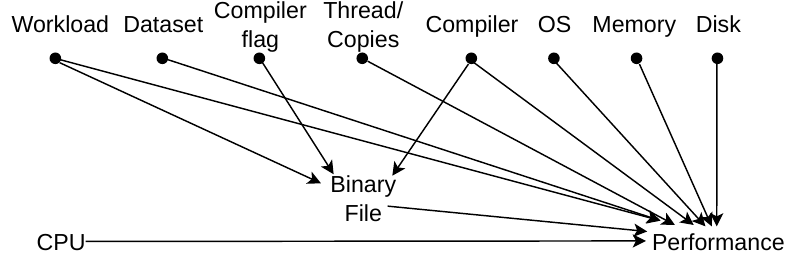}
\caption{
Causal graph of the true CPU effect.} 
\label{fig-cpu-effect}
\end{figure}

Figure~\ref{fig-cpu-eva-causal} shows the resulting causal graph. Crucially, with the exception of compiler flags, all other essential components manifest a dual impact on the system performance: they affect performance directly, whilst simultaneously modulating it indirectly via CPU execution state (The CPU node in Figure~\ref{fig-cpu-eva-causal} and Figure~\ref{fig-cpu-effect}, represents the CPU's behavior in scheduling and executing the dynamic instruction stream). To illustrate, the workload (source code) and the dataset (input data) intrinsically determine algorithmic time complexity, while they shape the behavior of the
dynamic instruction stream; copies/threads (the number of processes/threads) define execution granularity in the system, 
thereby influencing both overall execution cost and CPU scheduling/execution behavior;
OS and compiler infrastructures (e.g., schedulers and linkers/runtime support) inject performance overheads while also reshaping the execution flow; and memory/disk specifications inherently govern system access latencies, dictating the CPU’s architectural resource
accesses. Compiler flags, conversely, exert no direct performance impact; their causal influence is mediated by the binary file synthesized through the interplay of the workload source code, the compiler infrastructure, and the specific compiler flag. 
The binary file then serves as the static basis for the dynamic instruction stream executed by the CPU.

To infer the true CPU effect, we formalize the isolated causal path as follows: 
$CPU \rightarrow Performance$
Yet, the graph explicitly exposes that empirical measurements are heavily convoluted by confounding bias. This interference propagates through non-causal backdoor paths 
($CPU \leftarrow X \rightarrow Performance$)
, where $X$ represents the \textit{confounder set} including binary file and other components except compiler flags. Ultimately, CPU Evaluatology provides a rigorous structural explanation for the pervasive confounding bias in contemporary CPU benchmarking.

\subsection{Inferring the true CPU effect}\label{step-three}

\begin{figure*}[ht]
\centering
\includegraphics[scale=0.42]{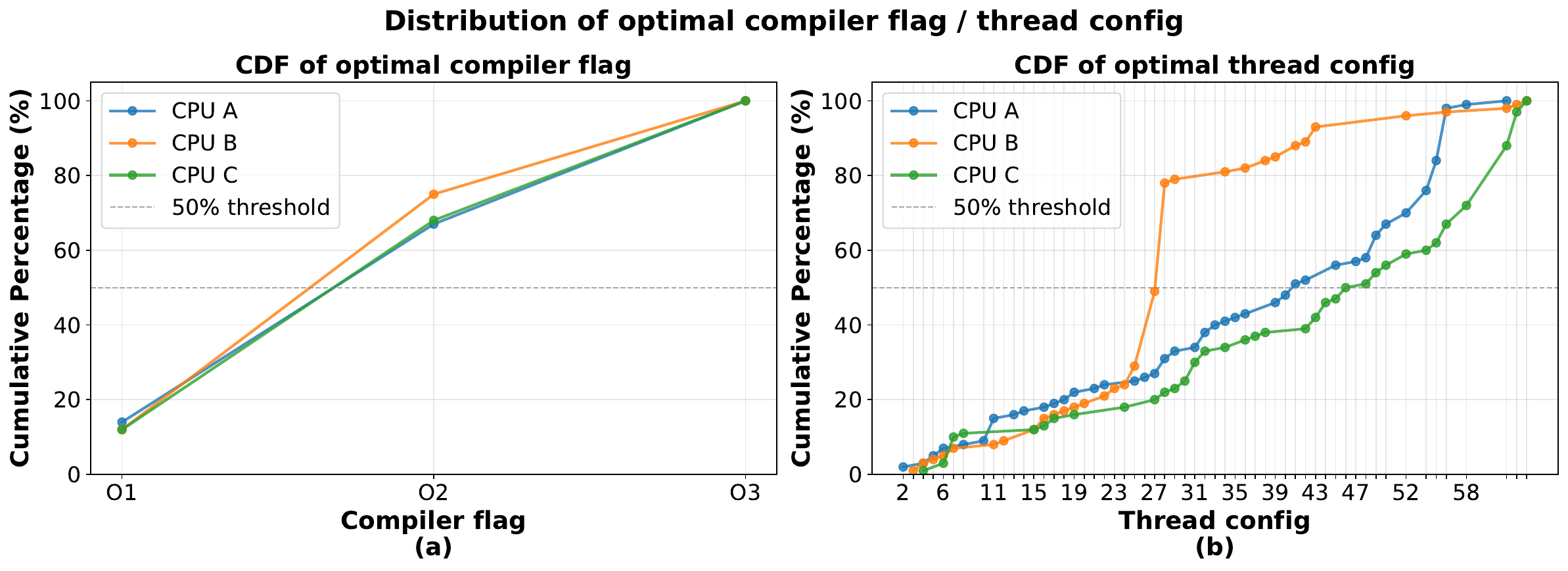}
\caption{
Distribution of compiler flags and threads measured across three CPUs from Table~\ref{Machine-Inf}, utilizing 10 datasets from the CPUOracle suite under the \texttt{SPECfpspeed} workloads.}  
\label{fig-opt-config}
\end{figure*}

Contemporary benchmark-centric paradigms infer CPU performance by aggregating static workloads while anchoring other essential components to empirical, deterministic configurations. This mechanism effectively neutralizes the $CPU \leftarrow Workload \rightarrow Performance$ backdoor path. However, the successful obstruction of other backdoor paths hinges entirely on the tenuous assumption of deterministic configuration distributions. While infrastructural substrates—namely compilers, OS, memory, and disks—can be viably modeled as temporally static, as shown in Figure~\ref{fig-opt-config}, the configuration distributions for dataset and copies/threads fundamentally defy such deterministic constraints. Thus, current benchmark-centric methodologies inherently fail to exhaustively seal all backdoor paths, precipitating systemic confounding bias.

In contrast, rigorous formalisms like DoE and RCTs successfully derive the true CPU effect by exhaustively traversing all confounding permutations to eliminate backdoor paths. Yet, this exhaustive exploration precipitates an exponential combinatorial explosion of the system configuration space, imposing computationally prohibitive and practically intractable overheads.

CPU Evaluatology circumvents this dilemma in a highly cost-effective by using random-sampling based on causal model outcomes. 
Recognizing that workloads persistently monopolize the variance contribution in ANOVA (Table~\ref{ANOVA-F-test-SPEC} and  Table~\ref{ANOVA-F-test-SPEC-EEC}), function as canonical proxies for execution tasks, and are requisite for equitable benchmarking comparisons, we purposefully architect a workload-stratified random sampling methodology.

\subsection{The Metrics}
We formalize the metrics utilized to infer the true CPU effect on the system performance. These metrics are divided into two classes: population-level parameters derived via exhaustive traversal of the configuration space, and sample-based estimators derived through stochastic sampling of the configuration space.

First, we quantify the performance discrepancy between the evaluated CPU ($c_i$) and the reference one ($c_{ref}$), denoted as \textbf{$\Delta E$} by analyzing their respective system performance under varying other components configurations ($X$). For each configuration $x_k$ within the configuration space, the differential effect is captured by the term $OE(c_i \mid x_k) - OE(c_{ref} \mid x_k)$. Since this pair of experiments uses a uniform configuration $x_k$, the resulting difference is solely attributable to the true functional divergence between $c_i$ and $c_{ref}$.  Then, $E(c_i) = \mathbb{E}_{X \sim P(X)} \left[ OE(c_i \mid X) \right]$ and $E(c_{ref}) = \mathbb{E}_{X \sim P(X)} \left[ OE(c_{ref} \mid X) \right]$, $P(X)$ is the configuration distribution.  Consequently, the CPU effect difference is formalized as Equation~\ref{EQ_CPU_delta}.
\begin{equation}
\Delta E = E(c_i) - E(c_{ref})\label{EQ_CPU_delta}
\end{equation}

Second, we employ the Confidence Interval (CI) of $\Delta E$ as a statistical metric to evaluate the accuracy of the CPU effect estimation. Specifically, for a given confidence level $1-\alpha$, we construct the interval as in Equation~\ref{EQ_CPU_interval}.
\begin{equation}
CI = \left[ \Delta \bar{E} - t_{\alpha/2,n-1}\frac{s}{\sqrt{n}}, \Delta \bar{E} + t_{\alpha/2,n-1}\frac{s}{\sqrt{n}} \right],
\label{EQ_CPU_interval}
\end{equation}
where $\Delta \bar{E}$ is the sample mean of the differences, $s$ is the sample standard deviation, $n$ is the number of experimental samples, and $t_{\alpha/2,n-1}$ represents the critical value of the $t$-distribution with $n-1$ degrees of freedom at the significance level $\alpha/2$. 
If the estimated value falls within this interval, it demonstrates that the estimation achieves the required accuracy ($1-\alpha$).

\section{Verifying the Correctness of CPU Evaluatology}\label{Sec_EffectOracle}

\subsection{Establishing Test Oracles}\label{step-four}

To verify the correctness of CPU Evaluatology, we first formalize two \textbf{Test Oracles}, and then we instantiate them into a tool, which we name CPUOracle.

\textbf{Test Oracle 1: The effect of the identical CPU on the same system should be a constant.} That is to say, when the system passes the completeness testing, for a metric measuring the effect of the identical CPU on the same system, the delta of the performance numbers of the same CPU should be negligible.

However, as diagnosed in Section~\ref{pitfalls-spec}, the SPEC CPU methodology relies on static workload configurations, fundamentally incapable of eliminating confounding bias. Consequently, they produce wildly varied performance numbers for the identical CPU. Instead, the very small error value (0.1501\%) reported in Table~\ref{ANOVA-F-test-SPEC-EEC} suggests that CPU Evaluatology has effectively eliminated confounding bias.

\textbf{Test Oracle 2: The cross-CPU performance difference under the same and consistent compiler flags should remain invariant.}
The CPU effect causal graph (Figure~\ref{fig-cpu-effect}) demonstrates that compiler flags lack a direct causal path to performance; rather, their causal influence (causal path) is mediated by the synthesized binary files ($Compiler\ flag\rightarrow Binary\ file\rightarrow Performace$). By applying the backdoor criterion to traverse all confounding components, we explore the configuration distributions of all system components except the compiler flag to sever the backdoor paths. Thus, fixing the compiler flag  marginalizes and controls the Binary file. This conditioning blocks the ($Compiler \ flag\rightarrow Binary\ file\rightarrow Performance$)
path. As both Workload and Compiler are already controlled, the path to Performance  
($Compiler\ flag  \rightarrow Binary\ file\leftarrow Workload/Compiler\rightarrow Performance$) remains blocked. Consequently, during the derivation of the true CPU effect, controlling the confounder set $X$, including binary file and other components except compiler flags, isolates the Compiler flag from Performance. By applying Rule 1 (insertion/deletion of observations) of do-calculus~\cite{pearl2018book}, the true CPU effect is constant under a uniform compiler flag. Hence, the cross-CPU performance comparison under the same compiler flag is fundamentally an invariant.
This structural mechanism directly formulates Test Oracle 2 in CPU Evaluatology for rigorous cross-CPU performance comparison.

\subsection{CPUOracle suite: the Instantiation of Test Oracles in CPU Evaluatology}

Anchored by the test oracles in CPU Evaluatology, we instantiated and open-sourced the CPUOracle suite, a verification tool including 43 SPEC CPU2017 workloads alongside 10 custom micro-benchmarks. Moreover, CPUOracle suite applies our causal inference methodology by transmuting static benchmark workloads into configurable instantiation distributions across all system components (as outlined in Table~\ref{Oracle-suite}). Dictated by Test Oracle 2, we explicitly parameterize the compiler flags across three tiers (\texttt{-O1}, \texttt{-O2}, \texttt{-O3}) to test the inferred CPU effect.

\begin{table}[]
\caption{Construction of CPUOracle suite.}
 \resizebox{\columnwidth}{!}{
\begin{tabular}{lcccc}
\hline
\textbf{Benchmark} & \textbf{Workload} & \textbf{Dataset} & \textbf{Copies/Threads} & \textbf{Compiler Flag} \\ \hline
SPEC CPU2017       & 43                & configurable     & configurable            & 3                      \\
Micro-Bench         & 10                & configurable     & configurable            & 3                      \\ \hline
\end{tabular}
}
\label{Oracle-suite}
\end{table}

While the 43 canonical SPEC CPU2017 workloads remain structurally intact (\textbf{Note}: these require a commercial SPEC license and are excluded from our open-source repository), we massively augmented their input dataset space. By reverse-engineering the generation paradigms of the three official SPEC datasets, we synthesized an expanded corpus guaranteeing a minimum of 30 valid datasets per workload. Every synthetic dataset was subjected to rigorous empirical correctness validation against the official SPEC execution semantics prior to publication. This expansive dataset corpus is entirely open-sourced within the CPUOracle suite, complemented by automated scripting frameworks to generate copies/threads configurations.

Complementing SPEC workloads, we instantiated multi-threaded versions of 10 fundamental micro-benchmarks sour-ced from the open-source community and our handcrafted: AES~\cite{aes}, Compress~\cite{compress}, FIO~\cite{fio}, LU~\cite{LU}, matmul~\cite{matmul}, Run\_S-tream~\cite{Stream}, SHA256~\cite{sha256}, SHA256\_custom, SOR~\cite{SOR}, and Merge Sort. Each workload is implemented to expose fully tunable dataset scales and copies/threads granularities.

\subsection{Validating Test Oracle 2: Cross-CPU Architectural Invariance}\label{Test-Oracle2-Val}

In this subsection, we answer the following  question:
Does empirical data substantiate Test Oracle 2 (the invariance of cross-CPU comparison under a uniform compiler flag) across real-world workloads and distinct compilers?

Test Oracle 2 of CPU Evaluatology establishes a fundamental architectural invariant: the differences between the effect of distinct processors on the system performance should remain invariant under a uniform compiler flag.

To empirically validate this assertion, we deploy the CPUOracle suite to evaluate the cross-CPU performance differences among the three processors (Table~\ref{Machine-Inf}). We rigorously calculate the geometric mean of these performance ratios across the exhaustive configuration space formalized in Section~\ref{Sec_Exp_set}. Table~\ref{Oracle2-val} presents these ratios of the effect across the \texttt{-O1}, \texttt{-O2}, and \texttt{-O3} optimization tiers, utilizing both macroscopic real-world proxies (\texttt{SPECfpspeed}) and micro-benchmarks. 
To eliminate compiler-specific artifacts, Table~\ref{Oracle2-compiler} broadens the evaluation using two divergent toolchains (GCC9 and Clang14) on the micro-benchmarks. (Note: \texttt{SPECfpspeed} evaluations only utilize GCC9, as Clang14 encounters unresolved compilation failures on a subset of SPEC workloads.)

\begin{table}[]
\centering
\caption{Ratios of the effect of the three CPUs from Table~\ref{Machine-Inf} evaluated using the CPUOracle suite, and the validation of Test Oracle 2 in CPU Evaluatology.}
\resizebox{\columnwidth}{!}{
\begin{tabular}{llccc}
\hline
\multicolumn{1}{c}{\textbf{Workload}}  & \multicolumn{1}{c}{} & \textbf{CPU A / CPU B}         & \textbf{CPU A / CPU C} & \textbf{CPU B / CPU C}         \\ \hline
                                       & -O1                  & 1.0146                         & 0.8183                 & 0.8065                         \\
                                       & -O2                  & 1.0245 & 0.8524                 & 0.8320                         \\
\multirow{-3}{*}{\textbf{\texttt{SPECfpspeed}}} & -O3                  & 1.0152                         & 0.8348                 & 0.8223                         \\ \hline
                                       & -O1                  & 0.8595                         & 1.1477                 &  1.3354                        \\
                                       & -O2                  & 0.8593 & 1.1572                 & 1.3466 \\
\multirow{-3}{*}{\textbf{MicroBench}}  & -O3                  & 0.8610                        & 1.1697                 & 1.3586                         \\ \hline
\end{tabular}
}
\label{Oracle2-val}
\end{table}

\begin{table}[]
\centering
\caption{Validation of Test Oracle 2 among different compilers (gcc and clang).}
\resizebox{\columnwidth}{!}{
\begin{tabular}{llccc}
\hline
\multicolumn{1}{c}{\textbf{Compiler}}  & \multicolumn{1}{c}{} & \textbf{CPU A / CPU B}         & \textbf{CPU A / CPU C} & \textbf{CPU B / CPU C}         \\ \hline
                                       & -O1                  & 0.8564                        & 1.1533                 & 1.3467                         \\
                                       & -O2                  & 0.8568 & 1.1623                 & 1.3566                         \\
\multirow{-3}{*}{\textbf{Clang14}} & -O3                  & 0.8585                         & 1.1654                 & 1.3575                         \\ \hline
                                       & -O1                  & 0.8626                         & 1.1422                 &  1.3242                        \\
                                       & -O2                  & 0.8618 & 1.1520                 & 1.3367 \\
\multirow{-3}{*}{\textbf{GCC9}}  & -O3                  & 0.8635                        & 1.1739                 & 1.3596                         \\ \hline
\end{tabular}
}
\label{Oracle2-compiler}
\end{table}

From these comprehensive empirical observations, we distill three critical insights.

\textbf{Cross-CPU Architectural Invariance:} The empirical data rigorously substantiates Test Oracle 2. Across all configurations, the relative variance of the true effect of the three CPUs under a uniform compiler flag is tightly confined to a \textbf{4\% margin} across \texttt{O1}, \texttt{O2}, \texttt{O3}. This not only decisively proves that true hardware disparities are fundamentally invariant under a uniform compiler flag, but also provides a \textbf{novel architectural insight}: the cross-CPU architectural invariance shows almost negligible differences \textit{across} compiler flags (\texttt{O1}, \texttt{O2}, \texttt{O3}).
    
\textbf{Generality Across Workloads and Compilers:} The validated architectural invariance demonstrates universal applicability. It holds consistently across starkly contrasting workload scales (macroscopic \texttt{SPECfpspeed} vs. micro-benchmarks) and remains robust across fundamentally different compiler infrastructures (GCC9 vs. Clang14). This confirms that Test Oracle 2 is fundamentally agnostic to system component specificities.
    
\textbf{Demystifying SPEC's False Variances:} These findings expose a critical flaw in traditional benchmarking paradigms. As previously demonstrated in Table~\ref{Mot-Cross-CPU}, SPEC CPU2017 evaluations report massive performance fluctuations and even conclusion inversions under a uniform compiler flag. Our validation proves that these deviations are not true architectural traits. Rather, they are purely symptomatic of unmitigated confounding bias—artifacts of uncontrolled system component configurations masking the true CPU capabilities. CPU Evaluatology successfully eliminates these confounders, revealing the underlying architectural truth.

\section{Evaluation}\label{TheEvaluationsection}

This section systematically evaluates CPU Evaluatology by using Hyper-Threading (HT) as a case study, verifying its accuracy, cost-efficiency, and essential role in inferring the true CPU effect, while demonstrating its ability to correct misleading conclusions from SPEC CPU and avoid the prohibitive costs of DoE and RCTs.

\subsection{Experimental Setup \& The Configuration Space}\label{Sec_Exp_set}

Table~\ref{Machine-Inf} summarizes the three processors evaluated. CPU A and CPU B are identical x86-based Intel Xeon Gold 5120T processors, differing only in HT activation, enabling us to infer the HT effect. CPU C is an ARM-based Kunpeng 920 processor serving as the cross-architecture counterpart. All machine component configurations adopt the default configuration shown in Table~\ref{Machine-Inf}.

To conduct an exhaustive baseline evaluation, we leveraged the CPUOracle suite to construct a massive benchmark component configuration space. Specifically, we swept across:
\begin{enumerate}
    \item \textbf{Workloads:} \textbf{10} \texttt{SPECfpspeed} workloads and \textbf{10} specialized micro-benchmarks.
    \item \textbf{Datasets:} \textbf{10} progressively scaled input datasets per workload, anchored by empirical heuristics and SPEC's official \texttt{test} set.
    \item \textbf{Threads:} \textbf{64} valid thread configurations, systematically scaling from 1 to 64 to capture the hardware topologies of the three CPUs. Crucially, to accommodate architectural execution constraints within the 654.roms\_s benchmark, thread configurations of 51, 53, 59, and 61 were purposefully excluded, yielding \textbf{60} valid configurations for \texttt{SPECfpspeed} workloads.
    \item \textbf{Compiler Flags:} \textbf{3} official optimization tiers (\texttt{-O1}, \texttt{-O2}, \texttt{-O3}).
\end{enumerate}

\subsection{Demystifying the Architectural Effect of Hyper-Threading}\label{ANOVA-HT}

To demonstrate how CPU Evaluatology rigorously infers system performance to a specific architectural component, we present a comprehensive case study: inferring the true effect of HT activation on the Intel Xeon Gold 5120T. We contrast CPU Evaluatology against the  SPEC CPU methodology and two general-purpose rigorous methodologies (DoE and RCTs).

\noindent\textbf{The Methodological Showdown.} 
We formulated the evaluation under three distinct paradigms utilizing the \texttt{SPECfpspe-}\\\texttt{ed} workloads:
\begin{enumerate}
    \item \textbf{SPEC CPU2017:}  Mandated by official SPEC rules, configurations are static. Thread counts are strictly locked to the CPU's hardware logical cores: \textbf{56 threads} when HT is enabled (CPU A) and \textbf{28 threads} when disabled (CPU B), evaluated solely on the \texttt{ref} dataset.
    \item \textbf{CPU Evaluatology \& DoE:} Using the CPUOracle suite, we generated a massive configuration space encompassing \textbf{10} workloads, \textbf{10} datasets, \textbf{3} compiler flags, and \textbf{60} thread configurations (excluding 4 incompatible 654.roms\_s configs). This yielded an exhaustive \textbf{36,000  configurations} across CPU A and CPU B. CPU Evaluatology infers the HT effect by comparing CPU A and CPU B under equivalent configurations, whereas the DoE mathematically integrates HT as a formal factor within its model equations. By solving the DoE model equations over this entire population, it established the mathematical ground truth.
    \item \textbf{Randomized Controlled Trials:} We designated CPU A as the \textbf{treatment group} and CPU B as the \textbf{control}, randomly assigning the 18,000 configurations in a 1:1 ratio to compute the Average Treatment Effect (ATE).
\end{enumerate}

\begin{table}[htbp]
  \centering
  \caption{Various methods were employed to infer the HT effect on Intel Xeon Gold 5120T using \texttt{SPECfpspeed} sub-suite.}
    \begin{tabular}{lc}\hline
    \textbf{Methodology} & \multicolumn{1}{l}{\textbf{CPU A / CPU B}}  \\\hline
     SPEC CPU2017   & 0.8994   \\
      CPU Evaluatology & 1.0181  
    \\
    DoE  &    1.0181 \\
    RCTs  &  1.0403  \\\hline
    \end{tabular}%
  \label{HT_off/HT_on}%
\end{table}%

\noindent\textbf{The Conclusion Inversion.} 
Table~\ref{HT_off/HT_on} exposes a catastrophic failure in traditional benchmarking.
The static SPEC methodology reports a deceptive 0.06\% performance degradation upon HT activation, dangerously misleading architects to conclude that this CPU heavily penalizes HT. 
In stark contrast, the rigorous traversal of the complete configuration space (CPU Evaluatology / DoE) reveals the unconfounded truth: HT inherently delivers a 1.81\% performance uplift. The RCTs similarly corroborate a net positive effect (+4.03\%). This drastic conclusion inversion proves that SPEC methodology fundamentally overfits to static configurations, failing to extract the true architectural effect.

\begin{figure}[ht]
\centering
\includegraphics[scale=0.5]{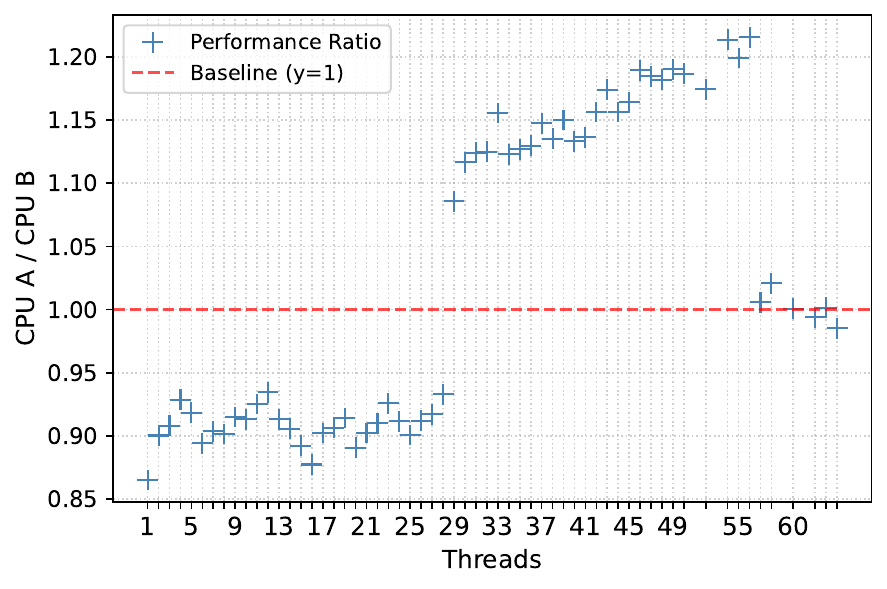}
\caption{
Fine-grained causal analysis of the HT effect with distinct thread-count clusters.} 
\label{fig-group-by-t}
\end{figure}

\noindent\textbf{Root Cause Analysis: Unmasking the Thread Confound-er.} 
Why did SPEC completely invert the architectural reality? Recognizing that HT is intimately coupled with thread-level parallelism, we stratified the 18,000 configuration measurements into 60 distinct thread-count clusters. As illustrated in Figure~\ref{fig-group-by-t}, the true HT effect exhibits a highly sensitive, piecewise correlation dictated entirely by thread scale.

\textbf{The Physical Bound Interval (threads$\in$[1, 28]):} When threads are constrained within the physical core capacity, HT introduces resource contention overhead, inflicting a severe performance penalty ranging from 6.54\% to 13.48\%. Within the physical bound interval, the performance penalty incurred by HT activation manifests as cyclical oscillations, periodically intensifying and receding with ascending thread counts. Mechanistically, when the thread count is bounded by the physical core capacity, concurrency demands are satisfied by dedicated physical cores. 
Therefore, in this scenario, the main effect of HT is not to increase core-level parallel capacity, but rather to introduce intra-core contention. This occurs when co-scheduled sibling threads compete for shared execution resources and L1/L2 caches, thereby incurring a performance penalty.

\textbf{The Transition Interval (threads$\in$[29, 56]):} When thre-ads populate the gap between physical and logical cores, HT unleashes its intended architectural potential, delivering substantial performance dividends from 8.57\% to 21.56\%. Within the transition interval, the performance dividends delivered by HT exhibit a consistent, overarching upward trajectory as the thread count scales. As the thread count exceeds the physical core capacity, the HT-disabled configuration must multiplex more runnable threads over fewer hardware execution contexts, increasing scheduling overhead and potentially amplifying synchronization costs. HT mitigates part of this pressure by exposing additional hardware thread contexts to the OS.

\textbf{The Saturation Interval (threads$\textgreater$56):} Beyond the logical core threshold, context-switching dominates, diminishing HT's true effect to marginal fluctuations (<2.5\%). Ultimately, once the thread count surpasses the logical core capacity, the system enters a state of severe oversubscription. Hardware threads fundamentally fail to absorb the concurrency, driving the OS into thrashing context switches and severe lock contention.

\textbf{Insight:} These empirical trends rigorously expose the fatal flaw of the SPEC CPU methodology. By artificially anchoring thread configurations to the logical core count (comparing 56 threads on CPU A vs. 28 threads on CPU B), SPEC 
embeds thread scaling as a confounder. It conflates the hardware HT effect with OS scheduling and thread contention artifacts. CPU Evaluatology systematically disentangles this confounding bias, restoring the true architectural evaluation.

\subsection{Methodological Showdown: Accuracy vs. Cost in Causal Inference}\label{Sec_Inferring_HT}

While Section~\ref{ANOVA-HT} established the mathematical ground truth (+1.81\% HT effect) by exhaustively traversing 36,000 configurations, such exhaustive evaluation is fundamentally intractable in real-world scenarios. Practical computer evaluation necessitates sampling. The critical question, therefore becomes, which methodology can accurately infer the true architectural effect with the minimum sampling cost?

\noindent\textbf{Evaluation Methodology.} 
To statistically evaluate the four methodologies (SPEC CPU, DoE, RCTs, and CPU Evaluatology), we adopted a rigorous Monte Carlo simulation framework. For each methodology, we extracted configuration samples, calculated the 99\% confidence interval (CI) for the HT effect, and repeated this process for 10,000 iterations. 
To ensure comparability with SPEC, all methodologies use the geometric mean of performance ratios, which is equivalent to averaging log-transformed deltas, providing a rigorous foundation for CI derivation. We evaluate the methodologies along two dimensions.
We benchmark the methodologies across two dimensions (detailed in Table~\ref{Experiment_Design}):
\begin{enumerate}
    \item \textbf{Accuracy:} The frequency (probability) with which the calculated 99\% CI successfully encapsulates the ground truth ratio (1.0181).
    \item \textbf{Cost:} The number of system configurations required per iteration.
\end{enumerate}

\begin{table}
\centering
\caption{Accuracy and cost comparison of CPU evaluation methodologies. Accuracy computed over 10,000 iterations.}
\begin{tabular}{lcl}
\hline
\textbf{Methodology}     & \textbf{Cost (Configs)} & \textbf{Accuracy} \\\hline
SPEC CPU2017  & 20                           & 0\%   \\\hline
$2^{k=5}r$ Factorial Design  & 32                           & 11.02\%    \\
$w2^{k=4}r$ Factorial Design & 160        & 73.58\%       \\ 
General Factorial Design & 36,000        & 100.00\%       \\ \hline
RCTs                      & 20                         & 0.03\%               \\
RCTs                      & 160                         & 0.06\%              \\
RCTs                      & 9,600                         & 1.10\%               \\ 
RCTs                      & 14,000                         & 98.86\%            \\ 
RCTs                      & 18,000                         & 99.01\%              \\\hline
CPU evaluatology                      & 10                         & 1.42\%             \\ 
CPU evaluatology                      & 60                         & 11.73\%             \\
CPU evaluatology
& 100                         & 94.68\%             \\
CPU evaluatology
& 110                         & 99.08\%             \\\hline
\end{tabular}
\label{Experiment_Design}
\end{table}

\noindent\textbf{Experimental Setup for Baselines.} 
Instead of static selections (SPEC), the rigorous baselines employ varied sampling paradigms. For the \textbf{DoE (Factorial Design)}, we constructed $2^kr$  fractional designs with $k=5$ factors and $k=4$ factors, binarizing factors into high/low levels with $r=3$ replicates. For $k=5$ factors (HT, workload, dataset, compiler, threads), there are a total of $2^5=32$ sampled configurations. For $k=4$ factors, adopting a stratification paradigm analogous to CPU Evaluatology, we designate the \textit{10 workloads} as individual strata and perform a $2^kr$ factorial design on the 4 remaining factors (termed as $w2^{k=4}r$ fractional design). This yields a total of $10\times 2^4=160$ sampled configurations. For \textbf{RCTs}, we employed completely randomized, without-replacement assignment between the treatment (CPU A with HT-enabled) and control (CPU B with HT-disabled) groups. For \textbf{CPU Evaluatology}, we applied our proposed stratified random sampling (using workloads as strata) to pair equivalent configurations across CPUs.

\noindent\textbf{Critical Insights \& Results.} 
Table~\ref{Experiment_Design} reveals the striking disparity in sampling efficiency among the methodologies. We summarize the findings into four pivotal insights.

\textbf{SPEC CPU methodology is Fundamentally Biased:} Constrained by rigidly static configurations, SPEC completely fails to neutralize confounding biases. No matter how many iterations are run, its estimation inherently misses the ground truth, rendering it incapable of causal inference.
    
\textbf{DoE Suffers from Severe Undersampling:} The $2^kr$ full factorial designs within the DoE methodologies exhibited unacceptably poor accuracy (73.58\% for 160 configurations, plummeting to 11.02\% for 32 configurations). This failure stems from the severe information loss caused by forcibly binarizing a highly non-linear, multidimensional configuration space into merely 32 or 160 tested combinations. Consequently, conventional DoE methodologies are unsuited for the combinatorial explosion of computer systems.
    
\textbf{RCTs are Unscalably Expensive:} While RCTs eventually reach the 99\% accuracy target, they demand a staggering cost of 14,000 configurations—almost equivalent to exhaustively testing the entire population. In complex systems, purely random assignment without structural awareness yields massive variance, requiring brute-force scaling to achieve statistical confidence.
    
\textbf{CPU Evaluatology Achieves Orders-of-Magnitude Efficiency:} By leveraging domain-aware stratified sampling and strict configuration pairing, CPU Evaluatology successfully hits the 99\% accuracy target requiring merely \textbf{110 configurations}. This shrinks the causal estimation cost by \textbf{over two orders of magnitude} compared to general factorial design in DoE \& RCTs. Furthermore, even if constrained to an ultra-low budget equivalent to SPEC (e.g., 10 configurations), CPU Evaluatology delivers an unbiased architectural estimation free from the confounding artifacts that plague static benchmarks.

\textbf{Summary.} CPU Evaluatology is the only practically viable methodology that shatters the dilemma between prohibitive time costs (RCTs/DoE) and accuracy degradation (SPEC), establishing a highly cost-effective standard for architectural causal inference.

\section{Conclusion}\label{Sec_Conclusion}
This paper formulates CPU performance evaluation as a causal inference problem under confounding system interactions and proposes CPU Evaluatology, a general-purpose methodology combining completeness testing, structural causal modeling, and sampling for cost-effective effect inference. We introduce architectural invariants as test oracles for correctness and show through controlled experiments that existing approaches (DoE, RCTs, SPEC CPU2017) are either inefficient or unreliable. Although instantiated for CPU evaluation, CPU Evaluatology is general and can extend to other system components.



\bibliographystyle{plain}
\bibliography{refs}

\end{document}